\documentclass{PoS}
\usepackage{braket}
\renewcommand{\d}{\ensuremath{\mathrm{d}}}

\newcommand{\gf}{\ensuremath{\mathrm{gf}}}
\newcommand{\phys}{\ensuremath{\mathrm{phys}}}
\newcommand{\YM}{\ensuremath{\mathrm{YM}}}
\newcommand{\e}{\ensuremath{\mathrm{e}}}
\newcommand{\p}{\partial}

\newcommand{\GZ}{\ensuremath{\mathrm{GZ}}}

\title{A candidate for the scalar glueball operator within the Gribov-Zwanziger  framework }

\ShortTitle{Glueballs within the GZ framework}

\author{\speaker{Nele Vandersickel}, David Dudal, Henri Verschelde \\
        Ghent University, Department of Mathematical Physics and Astronomy \\
        Krijgslaan 281-S9, 9000 Gent,Belgium\\
        E-mail: \email{nele.vandersickel@ugent.be} ,\email{ david.dudal@ugent.be}, \email{henri.verschelde@ugent.be }}

\author{Silvio P. Sorella\\
        Departamento de F\'{\i }sica Te\'{o}rica, Instituto de F\'{\i }sica, UERJ - Universidade do Estado do Rio de Janeiro\\
        Rua S\~{a}o Francisco Xavier 524, 20550-013 Maracan\~{a}, Rio de Janeiro, Brasil\\
        E-mail: \email{sorella@uerj.br}
}

\abstract{This proceeding gives an overview of the renormalization of $F^2_{\mu\nu}$ using  the Faddeev-Popov action and the more complicate Gribov-Zwanziger action, which deals with Gribov copies. We show that using the Faddeev-Popov action, $F^2_{\mu\nu}$ mixes with other $d=4$ operators. However, due to the BRST invariance of the action, this mixing is not relevant at the level of the correlator, $\Braket {F^2_{\mu\nu}(x) F^2_{\alpha \beta}(y)}$. In contrast, when turning to the Gribov-Zwanziger action, the mixing of $F^2_{\mu\nu}$ with other $d=4$ operator does have consequences at the level of the correlator. This is due to the breaking of the BRST. We then present a possible candidate for a physical operator in the Gribov-Zwanziger framework.}

\FullConference{International Workshop on QCD Green's Functions, Confinement, and Phenomenology - QCD-TNT09\\
		 September 07 - 11 2009\\
		 ECT Trento, Italy}

\begin{document}

\section{Introduction}
Quantum Chromodynamics (QCD) is the theory of interactions between quarks and gluons. As is well known, this theory is asymptotically free at high energies, while at low energies, confinement sets in. There is still no good understanding of confinement as non-perturbative aspects start to play an important role at the relevant energy scale. Therefore, it is of high interest to study pure gluonic particles where only gauge fields play a role. These pure gluonic particles are called glueballs.\\
\\
So far, there has been 20 years of intensive experimental search towards glueballs, see \cite{Crede:2008vw} for a complete overview on the experimental status. Unfortunately, no definite answer to the question whether a glueball has been observed or not can be given. Although, in the scalar sector, there is no question that more states have been found than can be accommodated by a single meson nonet. Hopefully, the future will provide us will more data on this as there are many new experiments planned, e.g.~the $\overline{\mathrm{P}}$ANDA Experiment at GSI in Germany \cite{Bettoni:2005ut}, BES III at BEPCII in Bejing \cite{Chanowitz:2006wf}, the GlueX Experiment at Jefferson Laboratory in the USA \cite{Carman:2005ps}, ALICE at CERN \cite{Alessandro:2006yt}.\\
As no clear experimental data is available on glueballs, lattice QCD has become very important for extracting relevant information about glueballs. In the quenched approximation, the state of the art is currently given by \cite{Chen:2005mg, Morningstar:1999rf}. For example, the masses of the three lowest lying glueballs are given by
\begin{eqnarray}
\mbox{scalar glueball}: 0^{++} &=& 1.710 \mbox{GeV}/\mbox{c}^2 \;, \nonumber\\
\mbox{tensor glueball}: 2^{++} &=& 2.390 \mbox{GeV}/\mbox{c}^2 \;,\nonumber\\
\mbox{pseudo scalar}: 0^{+-} &=& 2.560 \mbox{GeV}/\mbox{c}^2 \;.
\end{eqnarray}
On the theoretical side, various models and methods have been developed to calculate properties and masses of the glueballs. Historically, the first model to study glueballs is called the MIT bag model, and was applied to glueballs in 1976, see \cite{Jaffe:1975fd}. After this, many other models have been constructed, often phenomenological or with at least some input from experimental data. For an extensive overview, see \cite{Mathieu:2008me}. \\
\\
A very fundamental way to investigate bound states of gluons, is to study the correlator\\
$\braket{F_{\mu \nu}^2(x) F^2_{\alpha \beta}(y)}$, where $F^2_{\mu\nu}$ is classically the gauge field strength. In the next sections, we shall study this correlator using first the Faddeev-Popov action, and then the much more complicated Gribov-Zwanziger framework. In both theories, we shall show that one can construct a renormalization group invariant containing $F^2_{\mu\nu}$. However, the Yang-Mills gauge theory obeys a BRST symmetry $s$, while the Gribov-Zwanziger action breaks this BRST symmetry. This has big consequences for the correlator, as we shall show in the next sections.

\section{The construction of the glueball operator in the Yang-Mills action}
\subsection{The Yang-Mills action and the BRST symmetry}
When turning to the quantumlevel, one can overcome the problem of integrating over zero modes in the Yang-Mills action $\int \d^4 x \frac{1}{4} F_{\mu\nu}^2$, by introducing a gauge fixing. When choosing the Landau gauge, $\p_\mu A_\mu^a =0$, the partition function is given by
\begin{equation}\label{part}
Z_{FP} = \int [dA]\delta(\p A) \det M^{ ab} e^{-\frac{1}{4}\int \d^4 x F_{\mu\nu}^a F_{\mu\nu}^a}  \;,
\end{equation}
with
\begin{equation}
M^{ab} = -\p_\mu (\p_\mu \delta^{ab} - g f^{abc} A_\mu^c) \;,
\end{equation}
the Faddeev-Popov operator. We can lift the determinant into the action, by introducing the ghost fields $c$ and $\overline c$,
\begin{equation}
Z_{FP} = \int [dA][dc][d \overline c][d b] e^{-S_{FP}} \;,
\end{equation}
with
\begin{equation}
S_{FP} = \frac{1}{4}\int \d^4 x F_{\mu\nu}^a F_{\mu\nu}^a + \int \d^{4}x\;\left( b^{a}\partial_\mu A_\mu^{a}+\overline{c}^{a}\partial _{\mu } D_{\mu }^{ab}c^b \right)\;.
\end{equation}
We have also introduced the auxiliary $b$-field, which implements the Landau gauge.\\
\\
The Faddeev-Popov action is invariant under the following nilpotent BRST transformation,
\begin{equation}
s S_{FP} = 0 \;,\hspace{3cm} s^2 =0\;,
\end{equation}
with
\begin{equation}\label{s}
    sA_{\mu }^{a} =-\left( D_{\mu }c\right) ^{a}\;,    \hspace{0.5cm}
    sc^{a} =\frac{1}{2}gf^{abc}c^{b}c^{c}\;, \hspace{0.5cm}
    s\overline{c}^{a} =b^{a}\;, \hspace{0.5cm}
    sb^{a}=0\;.
\end{equation}
This BRST symmetry is of paramount importance as it is at the origin of the Slavnov-Taylor identity, which enables us to prove the renormalizability of the Yang-Mills action \cite{Piguet:1995er,Becchi:1975nq}.\\
\\
Another important property is that the BRST charge allows us to to define the sub-space of the physical states:
\begin{equation}\label{def}
s O_{\phys} = 0 \;,
\end{equation}
whereby $O_{\phys} + s X$ is in the same class of operators as $O_{\phys}$, since $s^2 X =0$ trivially. The BRST symmetry $s$ also allows us to establish the unitarity the $S$ matrix at the end \cite{Becchi:1974xu,Kugo:1979gm}.

\subsection{The construction of a physical operator $\mathcal R$}
Now, we can construct a scalar glueball operator containing $ F^2_{\mu\nu}$ which is renormalizable \cite{Dudal:2008tg}. For this, we introduce a dimensionless source $q$ to couple $F^2_{\mu\nu}$ to the Faddeev-Popov action $S_{FP}$. Naively, one would expect the following action to be renormalizable,
\begin{equation}\label{naief}
S_{FP} + \int \d^4 x q(x) F^2_{\mu\nu}(x) \;.
\end{equation}
However, $ F^2_{\mu\nu}$ will mix with other dimension four operators and therefore the action (\ref{naief}) is not renormalizable. \\
\\
There exist three types of dimension four operators. Firstly, we have the gauge invariant operators $\mathcal F$. These are constructed from the field strength $F_{\mu\nu}^a$ and the covariant derivative $D_{\mu}^{ab}$. These are BRST closed but not exact, e.g.~$F_{\mu\nu}^2$. Secondly, we have the gauge invariant exact operators $\mathcal E$, e.g.~$s(\overline{c}^a \p_\mu A_\mu^a)$, with $s$ the BRST variation (\ref{s}). Finally, one can construct ``Equation of motion terms'' $\mathcal H$, e.g.~$A_\mu^a \frac{\delta S}{\delta A_\mu^a}$, which vanish upon imposing the equation of motion.\\
 \\
Now, one can intuitively easily understand that these 3 different classes will mix in a certain way \cite{Collins:1984xc,Collins:1994ee,Henneaux:1993jn}. Firstly, a $\mathcal E$ operator cannot mix with a gauge invariant operator $\mathcal F$. Indeed, as
\begin{equation}
\Braket{\mathcal E_{0}}= \Braket{s(...)} = 0 \;,
\end{equation}
and suppose $\mathcal E$ mixes with the two other classes of operators,
\begin{equation}
\Braket{ \mathcal E_{0} }= a\Braket{ \mathcal F } + b\underbrace{\Braket{ \mathcal E}}_{{= 0}} + c \underbrace{\Braket{ \mathcal H } }_{{=0}} \;,
\end{equation}
we know that the mixing coefficient $a$ has to be equal to zero. Secondly, a $\mathcal H$ operator cannot mix with $\mathcal F$ and $\mathcal E$ as $\mathcal H$ will vanish upon using the equation of motion, while $\mathcal E$ as $\mathcal H$ will not. This can be translated in a upper triangular form of the mixing matrix,
\begin{eqnarray} \label{mixing}
 \left(
  \begin{array}{c}
    \mathcal F_0 \\
    \mathcal E_0 \\
    \mathcal H_0
  \end{array}
\right) &= & \left(
          \begin{array}{ccc}
            Z_{\mathcal F\mathcal F}& Z_{\mathcal F\mathcal E}  & Z_{\mathcal F \mathcal E} \\
            0  &Z_{\mathcal E\mathcal E} & Z_{\mathcal E \mathcal H}   \\
           0 & 0& Z_{\mathcal H \mathcal H}          \end{array}
        \right)
        \left(
\begin{array}{c}
    \mathcal F \\
    \mathcal E \\
    \mathcal H
  \end{array}
\right) \;.
\end{eqnarray}
The arguments for this mixing behavior are very general, and therefore expression (\ref{mixing}) should be valid for every gauge theory which has a BRST symmetry $s$.\\
\\
With this knowledge, we can construct an improved action of (\ref{naief}) which is renormalizable, by introducing other dimension 4 operators. We shall couple a source to all of the following operators
\begin{eqnarray}\label{ope}
    \mathcal F_0 &=& \frac{1}{4}F^2_{\mu\nu} \;, \nonumber\\
    \mathcal E_0 &=& s(\overline{c}^a \p_\mu A_\mu^a)  \;, \nonumber\\
    \mathcal H_0 &=&  A_\mu^a\frac{\delta S_{FP}}{\delta A_\mu^a} \;,
\end{eqnarray}
and analyze the renormalization following the formalism of algebraic renormalization \cite{Piguet:1995er}. Details of the calculations can be found in \cite{Dudal:2008tg}. We have proven that the new action including the three operators in expression (\ref{ope}) is renormalizable, whereby the mixing matrix is given by
\begin{eqnarray}\label{echtmix}
 \left(
  \begin{array}{c}
    \mathcal F_0 \\
    \mathcal E_0 \\
    \mathcal H_0
  \end{array}
\right) &= & \left(
          \begin{array}{ccc}
            Z_{qq}^{-1}& -Z_{Jq}Z_{qq}^{-1}  &-Z_{Jq}Z_{qq}^{-1} \\
            0  &1 & 0   \\
           0 & 0& 1          \end{array}
        \right)
        \left(
\begin{array}{c}
    \mathcal F \\
    \mathcal E \\
    \mathcal H
  \end{array}
\right) \;.
\end{eqnarray}
Notice the upper triangular form predicted in expression (\ref{mixing}), which is now clearly proven. We also see that $\mathcal E$ does not mix with $\mathcal H$, which is logical as $\mathcal E$ is a sum of two ``equation of motion'' terms and behaves like $\mathcal H$.\\
\\
Now that we have found the mixing matrix, we shall completely fix the entries of this matrix following the lines of \cite{Brown:1979pq}. Starting with the following most general $(n + 2m + r)$-point functions
\begin{eqnarray}\label{green}
\mathcal G^{n + 2m + r} &=& \Braket{A(x_1)\ldots A(x_n) c(y_1)\ldots c(y_m ) \overline c(\hat y_1) \overline c (\hat y_m) b(z_1)\ldots b(z_r)  }\nonumber\\
&=& \int [d \Phi] A(x_1)\ldots A(x_n) c(y_1)\ldots c(y_m ) \overline c(\hat y_1) \overline c (\hat y_m) b(z_1)\ldots b(z_r)  \e^{- S} \;,
\end{eqnarray}
it is possible to determine the mixing matrix (\ref{echtmix}) to all orders by using the fact that $\frac{\d \mathcal G^{n + 2m + r} }{\d g^2}$ must be completely finite. We have found that \cite{Dudal:2008tg}
\begin{eqnarray}\label{echtmixbepaald}
 \left(
  \begin{array}{c}
    \mathcal F_0 \\
    \mathcal E_0 \\
    \mathcal H_0
  \end{array}
\right) &= & \left(
          \begin{array}{ccc}
            1 -\frac{\beta / g^2}{\epsilon}& -\frac{2 \gamma_c}{\epsilon}  &-\frac{ 2 \gamma_c}{\epsilon} \\
            0  &1 & 0   \\
           0 & 0& 1          \end{array}
        \right)
        \left(
\begin{array}{c}
    \mathcal F \\
    \mathcal E \\
    \mathcal H
  \end{array}
\right) \;,
\end{eqnarray}
whereby $\beta$ is the $\beta$-function and $\gamma_c$ is the anomalous dimension of the ghost. Let us mention that this latter technique of determining the mixing matrix to all orders is only possible for the operator mixing matrix of the operator $F^2_{\mu\nu}$. This would not work for other operators e.g.~$F^4, F_{\mu \alpha} F_{\nu \alpha} - \frac{1}{d} \delta_{\mu\nu} F^2,$ etc. This is related to the fact that deriving the Green's functions (\ref{green}) w.r.t.~$g^2$ brings down the operator of interest $F^2_{\mu\nu} = \mathcal F$ from $\e^{-S}$.\\
\\
The final step is to determine a renormalization group invariant from (\ref{echtmixbepaald}). This can be done easily and we have found,
\begin{eqnarray}\label{2.18}
O_\phys = \mathcal R &=&  \frac{\beta(g^2)}{g^2} \mathcal F - 2 \gamma_c(g^2) \mathcal E- 2\gamma_c(g^2)\mathcal H  \;.
 \end{eqnarray}
This means that instead of $\Braket{\mathcal F(x) \mathcal F(y)}$, we should actually investigate the quantum version   \\
$\Braket{\mathcal R(x) \mathcal R(y)}$. We can make a few observations. Firstly, we see that this renormalization group invariant $\mathcal R$ coincides with the trace anomaly $\Theta_\mu^\mu$, which is expected as
$\Theta_\mu^\mu$ is also a $d=4$ renormalization group invariant \cite{Collins:1976yq}. Also, we recover the same mixing as in some older works in a slightly different formalism, see \cite{KlubergStern:1974rs, Joglekar:1975nu}.  Secondly, we can immediately set the term in $\mathcal H $ equal to zero as in the correlator we are working on-shell. Finally, the BRST symmetry plays an important role in the correlator. Due to $s$-invariance of the action, we find
\begin{eqnarray}
\Braket{O_{phys}(x) O_{phys}(y)} &=& \Braket{ \mathcal R(x) \mathcal R(y)}\nonumber\\
                                 &=&  \Braket{   \left[ \frac{ \beta} {g^2} F^2_{\mu\nu} + s(\ldots) \right](x)   \left[\frac{ \beta} {g^2} F^2_{\mu\nu} + s(\ldots) \right](y)} \nonumber\\
                                 &=& \Braket{ \left( \frac{ \beta} {g^2}\right)^2 F^2_{\mu\nu}(x) F^2_{\mu\nu}(y) + s(\ldots)}\nonumber\\
                                 &=& \left(\frac{ \beta} {g^2}\right)^2  \Braket{F^2_{\mu\nu}(x) F^2_{\mu\nu} (y)}\;,
 \end{eqnarray}
meaning that the mixing can be ignored in the Yang-Mills action at the level of the correlator due to the BRST symmetry.

\section{The construction of the glueball operator in the Gribov-Zwanziger action}
\subsection{The Gribov-Zwanziger action}
In 1977, Gribov discovered that the Landau gauge is  plagued by the existence of Gribov copies \cite{Gribov:1977wm}, i.e. there are still multiple gauge copies $A_\mu$ which all fulfill the Landau gauge condition. We can get partly rid of these copies by restricting the domain of integration in the Feynman path integral to the Gribov region $\Omega$ \cite{Zwanziger:1989mf},
\begin{equation}
\Omega = \left\{  A; \p A = 0 , - \p D > 0 \right\} \;,
\end{equation}
or thus by demanding the Faddeev-Popov operator to be strictly positive. The partition function (\ref{part}) is then translated into
\begin{equation}
Z_{H} = \int [dA]\delta(\p A) \det M^{ ab} e^{-\left(\frac{1}{4}\int \d^4 x F_{\mu\nu}^a F_{\mu\nu}^a + S_H \right)} \;,
\end{equation}
with $S_H$ a non-local term
\begin{equation}
S_H = \gamma^4 g^2 \int f^{abc} A^b_\mu (M^{-1})^{ad} f^{dec} A^e_\mu \;,
\end{equation}
the so-called horizon term \cite{Zwanziger:1989mf}. As a non-locality is difficult to handle, it is useful to recall that one can cast this action into a local form
\begin{eqnarray}
S_{\GZ} &=&S_{FP} + S_0+ S_{\gamma} \;,
\end{eqnarray}
with
\begin{eqnarray}
S_0 &=& \int \d^{4}x\left( \overline{\varphi}_\mu^{ac}M^{ab}\varphi^{bc}_\mu -\overline{\omega}_{\mu}^{ac} M^{ab} \omega_\mu^{bc} \right)  \;,\nonumber\\
S_{\gamma}&=& -\gamma ^{2}g\int\d^{4}x\left( f^{abc}(\varphi _{\mu }^{bc} +\overline{\varphi }_{\mu }^{bc} )A_{\mu }^{a} + \frac{4}{g}\left(N^{2}-1\right) \gamma^{2} \right) \;.
\end{eqnarray}
We have introduced a pair of complex conjugate bosonic fields $\left(\overline \varphi_\mu^{ac}, \varphi_\mu^{ac}\right)$ and a pair of Grassmann fields $\left( \overline \omega_\mu^{ac},\omega_\mu^{ac} \right)$. We recall that $S_\GZ$ is renormalizable to all orders \cite{Zwanziger:1992qr,Maggiore:1993wq,Dudal:2005na}, even in the presence of massless \cite{Gracey:2005cx,Gracey:2006dr} or massive quarks \cite{Ford:2009ar}.\\
 \\
The parameter $\gamma$ is called the Gribov parameter, has dimension of a mass and is not free but determined by
\begin{eqnarray}
\frac{\p \Gamma}{\p \gamma^2} &=& 0  \;,
\end{eqnarray}
with $\Gamma$ the quantum action defined as $\e^{-\Gamma} = \int [D\Phi] \e^{-S} $.\\
\\
Let us have a closer look at the BRST breaking of the Gribov-Zwanziger action. The new fields have the following BRST transformation:
\begin{equation}
s\varphi _{i}^{a} =\omega _{i}^{a}\;, \hspace{0.5cm} s\omega _{i}^{a}=0\;, \hspace{0.5cm} s\overline{\omega }_{i}^{a} =\overline{\varphi }_{i}^{a}\;,\hspace{0.5cm} s \overline{\varphi }_{i}^{a}=0\;. \nonumber\\
\end{equation}
One can check very easily that $S_{\GZ}$ is no longer invariant under the BRST symmetry $s$,
\begin{eqnarray}
 sS_\GZ = s \left(S_{\mathrm{YM}}+S_{gf} + S_0+ S_{\gamma}\right)  = s \left(S_\gamma \right) = g \gamma^2 \int \d^d x f^{abc} \left( A^a_{\mu} \omega^{bc}_\mu -
 \left(D_{\mu}^{am} c^m\right)\left( \overline{\varphi}^{bc}_\mu + \varphi^{bc}_{\mu}\right)  \right)\,.
\end{eqnarray}
This soft breaking has been studied in \cite{Dudal:2008sp}. Let us remark that despite this breaking, the Gribov-Zwanziger action is renormalizable, due to a rich set of Ward identities. Moreover, only two renormalization constants are needed, $Z_A$ and $Z_g$, just as in the Faddeev-Popov action. As a byproduct of this action, we see that the gluons get ``confined'' by the horizon. Indeed, as the gluon propagator has complex poles, gluons cannot describe physical excitations. In addition, we also observe that the Fourier transform shows positivity violation \cite{Dudal:2008sp}. The question which rises now, is how to define physical operators as the definition (\ref{def}) is no longer possible here.\\
\\
Before giving a \textit{possible} answer to this question, let us first mention that it is possible to embed $S_\GZ$ into a ``larger'' action. For this, we replace $S_\gamma$ with $S_s = s(\ldots)$,
\begin{eqnarray*}
S_\GZ &=& S_{\YM} + S_{\gf} + S_0 + S_\gamma \\
&\updownarrow&\\
\Sigma_\GZ &=& S_{\YM} + S_{\gf} + S_0 + S_s
\end{eqnarray*}
We have introduced new sources in $S_s$, but in the end, we can set these sources to the right values so that $ \left. S_s \right|_{\phys} = S_\gamma$. This implies that we did not change the theory. The advantage is that we have an action which is BRST invariant again.

\subsection{The construction of a potentially physical operator $\mathcal R$}
With $\Sigma_\GZ$ we can construct a glueball operator which is renormalizable, completely analogous as in the Yang-Mills case, and in the end we can take the physical limit to the physical Gribov-Zwanziger action $S_\GZ$. This is a highly non-trivial task, see \cite{Dudal:2009zh} for all the details. However, the final outcome is very peculiar. We find exactly the same expression as in (\ref{2.18}),
\begin{eqnarray}\label{2.18bis}
O_\phys = \mathcal R' &=&  \frac{\beta(g^2)}{g^2} \mathcal F - 2 \gamma_c(g^2) \left.\mathcal E'\right|_\phys- 2\gamma_c(g^2) \left.\mathcal H'\right|_\phys \;,
\end{eqnarray}
but $\mathcal E'$ is no longer BRST exact:
 \begin{eqnarray}
  \left.\mathcal E '\right|_\phys &=& s(  \p_\mu  \overline c^a  A_\mu^a  +  \p_\mu \overline \omega_i^a  D_\mu^{ab} \varphi^b_i ) + \gamma ^{2}  D_\mu^{ab} \left( \varphi_\mu^{ba} + \overline \varphi_\mu^{ba} \right) + d (N^2 - 1) \gamma^4  \,, \nonumber\\
 \left.\mathcal H' \right|_\phys  &=& A_\mu^a \frac{\delta S_\GZ}{\delta A_\mu^a} \;.
 \end{eqnarray}
The fact that expression (\ref{2.18bis}) has exactly the same form as (\ref{2.18}) is closely related to the fact that the Gribov-Zwanziger action has only two independent renormalization constants, $Z_A$ and $Z_g$ just in the ordinary Yang-Mills action. \\
\\
This time, $\mathcal E'$ will not be trivial at the level of the correlator. Not only is $\mathcal E'$ no longer BRST invariant, also the action $S_\GZ$ breaks BRST softly. Therefore,
\begin{eqnarray}
\Braket{O_{phys}(x) O_{phys}(y)} &=& \Braket{ \mathcal R'(x) \mathcal R'(y)}\nonumber\\
                                 &=&  \Braket{   \left[ \frac{ \beta} {g^2} \mathcal F  -2 \gamma_c \mathcal E' \right](x)   \left[ \frac{ \beta} {g^2} \mathcal F  -2 \gamma_c \mathcal E' \right](y)} \nonumber\\
                                 &\not=& \left( \frac{ \beta} {g^2}\right)^2 \Braket{  \mathcal F (x) \mathcal F(y) }
\end{eqnarray}
and thus the correlator has extra parts in comparison with the standard glueball correlator\\ $\left( \frac{ \beta} {g^2}\right)^2  \Braket{  \mathcal F (x) \mathcal F(y) }$. In fact, Zwanziger \cite{Zwanziger:1989mf} has studies the pole structure of
\begin{eqnarray}
    \int \d^4 x \Braket{F^2_{\mu\nu} (x) F^2_{\mu\nu} (0)} \e^{ipx}
  \end{eqnarray}
at tree level and he has reported that this correlator displays a physical cut at $p^2 = -2 \gamma^2$, while it also displays unphysical cuts at $p^2 =\pm 4 i \gamma^2$. It would therefore be interesting to investigate the cut structure of the improved correlator $\Braket{ \mathcal R'(x) \mathcal R'(y)}$, to see whether it can be called physical or not.

\subsection{An extra word about contact terms}
There is perhaps one point which needs to be clarified w.r.t. contact terms. In \cite{Dudal:2008tg,Dudal:2009zh} we have always safely neglected contact terms as such terms are usually ignored \cite{Collins:1984xc}. The origin of such contact terms is the following. The correlator $\Braket{F^2_{\mu\nu}(x) F^2_{\alpha\beta}(y)}$ stems from deriving the partition function w.r.t.~$q(x)$ and $q(y)$ and then setting all sources equal to zero. One can imagine, having a pure source term like $\int d^4 q(x) \p^4 q(x) $ into the action, this gives rise to a contact term $\sim \p^4 \delta (x-y)$. However, as we are not interested in the space time point $x=y$, this term shall always be zero, since $\delta(0) = 0$ in dimensional regularization. Therefore, we have always neglected this class of terms. \\
\\
One can object now, as going to Fourierspace, this term is no longer irrelevant. Indeed, \begin{eqnarray}
\p^4 \delta (x-y) &=& \int \frac{d^4 k}{(2\pi)^4} k^4 \e^{i k(x-y)}
\end{eqnarray}
giving rise to a term $\sim k^4$. It is therefore possible that a term like this will be necessary to kill divergences of the form $\frac{k^4}{\epsilon}$, present in the Fourier transform of $\Braket{F^2_{\mu\nu}(x) F^2_{\alpha\beta}(y)}$.  As already said, when going back to $x$- space, a term like this is zero in dimensional regularization for $x \not= y$. Moreover, in Fourierspace, contact terms will only generate polynomial contributions in the momentum and/or external mass scales. Such terms are therefore irrelevant when investigating the pole and cut structure of the propagator in momentum space.\\
\\
In conclusion, contact terms are irrelevant for the calculation of the pole and cut structure of $\Braket{\mathcal R'(x) \mathcal R'(y)}$, as they do not contribute. They are however needed to ensure a completely finite correlator function, see \cite{Collins:1984xc,Gracey:2009mj}.

\section{Conclusion}
We have investigated the renormalization of $F^2_{\mu\nu}$ using the standard Faddeev-Popov action, as well as the more complicated Gribov-Zwanziger action, which takes into account Gribov copies.\\
 \\
Firstly, in the Yang-Mills case, we have found that $F^2_{\mu\nu}$ mixes with a BRST exact operator $\mathcal E = s(\overline c^a \p_\mu A^a_\mu)$. We have constructed a renormalization group invariant
\begin{equation}
\mathcal R = \frac{\beta(g^2)}{g^2} \mathcal F - 2 \gamma_c(g^2) \mathcal E \;.
\end{equation}
However, at the level of the correlator, this BRST exact operator does not play any role,
\begin{eqnarray}
\Braket{ \mathcal R(x) \mathcal R(y)}  &=& \left(\frac{ \beta(g^2)} {g^2}\right)^2  \Braket{F^2_{\mu\nu}(x) F^2_{\mu\nu} (y)}\;,
 \end{eqnarray}
 due to the $s$-invariance of the Faddeev-Popov action. \\
 \\
 Secondly, in the Gribov-Zwanziger case, something peculiar happens. The issue of the Gribov copies leads to a modification of the Faddeev-Popov formula. The BRST invariance of this modified Faddeev-Popov action turns out to be softly broken. After studying the renormalization of $F^2_{\mu\nu}$, we have found again that $F^2_{\mu\nu}$ mixes with
$ \mathcal E' = s(\overline c^a \p_\mu A^a_\mu + \p_\mu \overline \varphi_\nu^{ac} D_\mu^{ab} \varphi^{bc}_\nu) + \gamma^2 D_\mu^{ab} ( \varphi^{ba}_\mu + \overline \varphi^{ba}_\mu )  + d \left(N^{2}-1\right)  \gamma^4$. Notice that $\mathcal E'$ is no longer BRST exact. The renormalization group invariant has the same form as in the Yang-Mills case,
\begin{equation}
\mathcal R' = \frac{\beta(g^2)}{g^2} \mathcal F - 2 \gamma_c (g^2)\mathcal E'\;.
\end{equation}
However, now $\mathcal E'$ does not vanish at the level of the correlator, as the BRST is  softly broken in the Gribov-Zwanziger action.
\begin{eqnarray}
\Braket{ \mathcal R(x) \mathcal R(y)} & =& \left(\frac{ \beta(g^2)} {g^2}\right)^2  \Braket{\mathcal F(x) \mathcal F (y)}  - 2 \gamma_c \frac{ \beta} {g^2} \Braket{\mathcal F(x) \mathcal E(y)} \nonumber\\
 && - 2 \gamma_c \frac{ \beta} {g^2} \Braket{\mathcal E(x) F^2(y)} + 4 \gamma_c^2 \Braket{\mathcal E(x) \mathcal E(y)}\;.
 \end{eqnarray}
The possibility that the breaking of the BRST in the Gribov-Zwanziger action might be relevant in order to construct the local physical operators in the presence of the Gribov horizon looks very attractive. This will be investigated in future work.\\
\\
Finally, as a last remark, let us mention that this analysis is not affected by taking into account dimension two condensates into the theory. Therefore, we would find exactly the same results using the so-called Refined Gribov-Zwanziger framework, see \cite{Dudal:2007cw, Dudal:2008sp}.

\section*{Acknowledgments.}
 We are grateful to J.~A.~Gracey, V. Mathieu, and A. Quadri for useful discussions. We would like to thank to organizers of the ``International Workshop on QCD Green's Functions, Confinement, and Phenomenology'' for the kind invitation and good organization. We would also like to thank all the participants for interesting discussions. D.~Dudal and N.~Vandersickel are  supported by the Research Foundation-Flanders (FWO). The Conselho Nacional de
 Desenvolvimento Cient\'{\i}fico e Tecnol\'{o}gico (CNPq-Brazil), the Faperj,
 Funda{\c{c}}{\~{a}}o de Amparo {\`{a}} Pesquisa do Estado
 do Rio de Janeiro, the SR2-UERJ and the Coordena{\c{c}}{\~{a}}o de
 Aperfei{\c{c}}oamento de Pessoal de N{\'{\i}}vel Superior (CAPES),
 the CLAF, Centro Latino-Americano de F{\'\i}sica, are gratefully acknowledged for financial support.

\end{document}